\documentclass[aps,prb,amssymb,reprint,groupedaddress]{revtex4-2}
\usepackage{graphicx}
\usepackage{amsmath}

\graphicspath{ {~/LaTeX/PAPER-TTFFA/} }

\usepackage{dcolumn}% Align table columns on decimal point
\usepackage{bm}% bold math
\usepackage{textcomp}

\begin{document}
	
\def\W{cm$^{-1}~$}		% cm-1	

%\preprint{}

\title{Temperature induced Neutral to Ionic phase Transition of the charge transfer crystal Tetrathiafulvalene-Fluoranil}


\author{Elena Ferrari}
\author{Francesco Mezzadri}
\author{Matteo Masino}
\email[]{matteo.masino@unipr.it}
%\thanks{}
\affiliation{Dipartimento di Scienze Chimiche,
	della Vita e della Sostenibilit\`a Ambientale (SCVSA)
	and INSTM-UdR Parma, Universit\`a di Parma, Parco Area delle
	Scienze, IT-43124 Parma, Italy}

\date{\today}

\begin{abstract}

  The Temperature Induced Neutral to Ionic phase Transition (TI-NIT)
  is a rare phenomenon occurring in mixed stack charge transfer (CT) crystals
  made up of alternating $\pi$-electron Donor (D) and Acceptor (A) molecules.
  We were able to grow crystals of Tetrathiafulvalene-Fluoranil (TTF-FA),
  and to show that it undergoes TI-NIT like the prototype CT crystal TTF-Chloranil.
  We characterized both room and low T phases through IR and Raman spectroscopy and XRD,
  demonstrating that while TTF-FA is quasi-neutral at room T, its ionicity jumps from 0.15 to 0.7
  at low T, therefore crossing the Neutral-Ionic borderline. The transition, occuring around 150K, is first order, with large thermal hysteresis and accompanied by crystal cracking.
  In the high T phase D and A molecules lie on inversion center, i.e. the stacks are regular,
  whereas the low T phase is characterized by the loss of the inversion symmetry along the stack
  as the stacks are strongly dimerized and by the doubling of the unit cell.

\end{abstract}

%\keywords{}

\maketitle

\section{Introduction}
\label{s:intro}
Mixed stack (ms) charge transfer (CT) organic cocrystals may undergo a rare phase transition named Neutral to Ionic Transition (NIT)\cite{Torrance1981}.
These crystals contain planar $\pi-$electron donor (D) and acceptor (A) molecules that overlap their frontier orbitals forming DADADA stack chains, due to the CT interaction.
Depending on the ionicity ($\rho$), that is the average charge located on D and A, ms CT crystals can be Neutral (N) if $\rho<0.5$ or Ionic (I) otherwise.
In few cases, low temperatures or high pressures drive the system from a N ground state to a I one, originating a collective CT called NIT.
Due to Peierls instability, the NIT is always associated with stack dimerization, that breaks the inversion symmetry of the chain. \par
Since dimerized stacks are polar, the I phase is potentially ferroelectric, as in the case of the prototypical Tetrathiafulvalene-Chloranil (TTF-CA)\cite{Torrance1981, Torrance1981a}
that undergoes NIT at 81 K. Interestingly, TTF-CA ferroelectricity has an electronic origin,
as the polarization is dominated by intermolecular charge transfer\cite{Kobayashi2012}.
Further studies on TTF-CA revealed that the same transition can be also photoinduced \cite{Koshihara1990,Koshihara1999,Collet2003-science}.
Since the first discovery of the NIT, other exotic phenomena such as dielectric anomaly \cite{Mitani1984-soliton,Mitani1987,Okamoto1991}, soft modes\cite{Masino2003,Girlando2008},
quantum phase transitions\cite{Horiuchi2003-science,Kagawa2017} have been observed in correspondence with it.\par
For the above reasons, the mechanisms of the NIT are still being investigated, both experimentally and theoretically.
It is known that the NIT is characterized by the competition between two instabilities:
a valence one and a structural one. While the former is favored by 3D Coulomb interactions,
that induce  discontinuous NIT, the latter is due to electron-lattice phonon coupling,
that drives the stack to a continuous dimerization\cite{Painelli1988-ZeroT,Girlando1986-Regular}.
The role of these microscopic parameters has been demonstrated by comparing TTF-CA with other ms CT crystals undergoing NIT\cite{Delchiaro2017}.\par
In this context, the TTF-haloquinone series has been extensively studied by chemical modification of the component molecules\cite{Horiuchi2001-jacs,Horiuchi2006-ChemPhys}.
However, the elusive TTF-FA, first obtained in 1979 together with TTF-CA\cite{Mayerle1979}, has never been experimentally reproduced and characterized.
We were able to grow it as single crystals and to observe that it also undergoes NIT at low temperature.
Both phases were characterized by vibrational spectroscopy and XRD.

\section{Experimental}
%
%\subsection{Crystal growth}

Several attempts to grow TTF-FA crystals with slow methods always yielded a red-brown amorphous product.
Good quality single crystals were finally obtained by mixing saturated toluene solutions of the
two components in 1:1 stochiometric ratio and rapidly evaporating the solvent.
The crystals were unstable at room temperature due to FA sublimation and were stored at low temperature.\par
Single crystal X-rays diffraction was collected at RT by using graphite monochromatized Mo K$\alpha$ wavelength on a Bruker Smart diffractometer
equipped with an APEX II CCD detector.
Low temperature data were collected with a Bruker D8 Venture instrument, equipped with a Photon II CCD area detector
and micro-focused Mo K$\alpha$ radiation source. Temperature control was achieved with an Oxford cryostream system working in nitrogen flux.
Due to crystal breaking at the NIT transition, the low temperature data collection required the sample to be placed
into a glass capillary filled with Fomblin\textsuperscript{\textregistered}, a perfluorinated polymer that increases its viscosity as temperature is lowered.
In such a way crystal cracking was avoided without inhibiting the transition. 
%On the other hand the small crystal size together
%with the absorption ascribed to the glass walls the low temperature data quality was affected, 
%in particular at high sin$\theta$/$\lambda$.
The single crystals data reduction was carried out by using the SADABS program \cite{SHELX}
The software Shelxt was used for structures solution while refinement was carried out full-matrix by using the Shelxl program \cite{APEX2}.\par
The IR spectra were recorded with a Bruker IFS-66 FT-IR spectrometer coupled to a Hyperion 1000 IR microscope, on the $ac$ and on the $ab$ face
with the electric field vector polarized parallel or perpendicular to the stack direction $c$. The crystal faces were previously indexed by XRD.
%The measurements were performed in absorbance mode on the thinnest crystals, using KBr as a reference.
\par
The Raman spectra were recorded with a Renishaw 1000 spectrometer equipped with the appropriate edge filter and coupled
to a Leica M microscope using the 568.2 nm line of a Lexel Kr laser. Low frequency spectra have been measured using the 633 nm HeNe laser
and a Horiba LabRAM HR Evolution Raman spectrometer equipped with a ULF Bragg filter.
The laser power was always set below 0.1 mW to avoid sample heating.
Raman spectra were measured with the exciting and scattered light polarized parallel or perpendicular to the stack on both ac and ab crystal faces. \par
The crystals were cooled down to 80 K using a Linkam HFS 91 stage and were fixed to it to prevent sample disintegration during the phase transition.
In the case of Raman measurements, the crystals were pasted to the stage with a conductive silver paste,
while they were covered with a thin and flat KBr slide for transmission IR measurements.\par
Standard DFT computational methods (B3LYP, 6-31G(d)) were exploited for the calculation of equilibrium geometry and vibrational frequencies
of the constituent molecules, in both the neutral and ionized state, using Gaussian 16 B.10\cite{gaussian16}.
The frequencies higher than 1000 \W were scaled by the factor 0.9613, as suggested in \cite{Merrick2007}. First principles calculations
following the approach described in \cite{Delchiaro2017} have been performed on the crystal structures to estimate
the tendency towards valence instability.

\section{Results and Discussion}

\subsection{Room temperature crystal structure}

\begin{figure}[hp]
	\centering
       	\includegraphics[width=\linewidth]{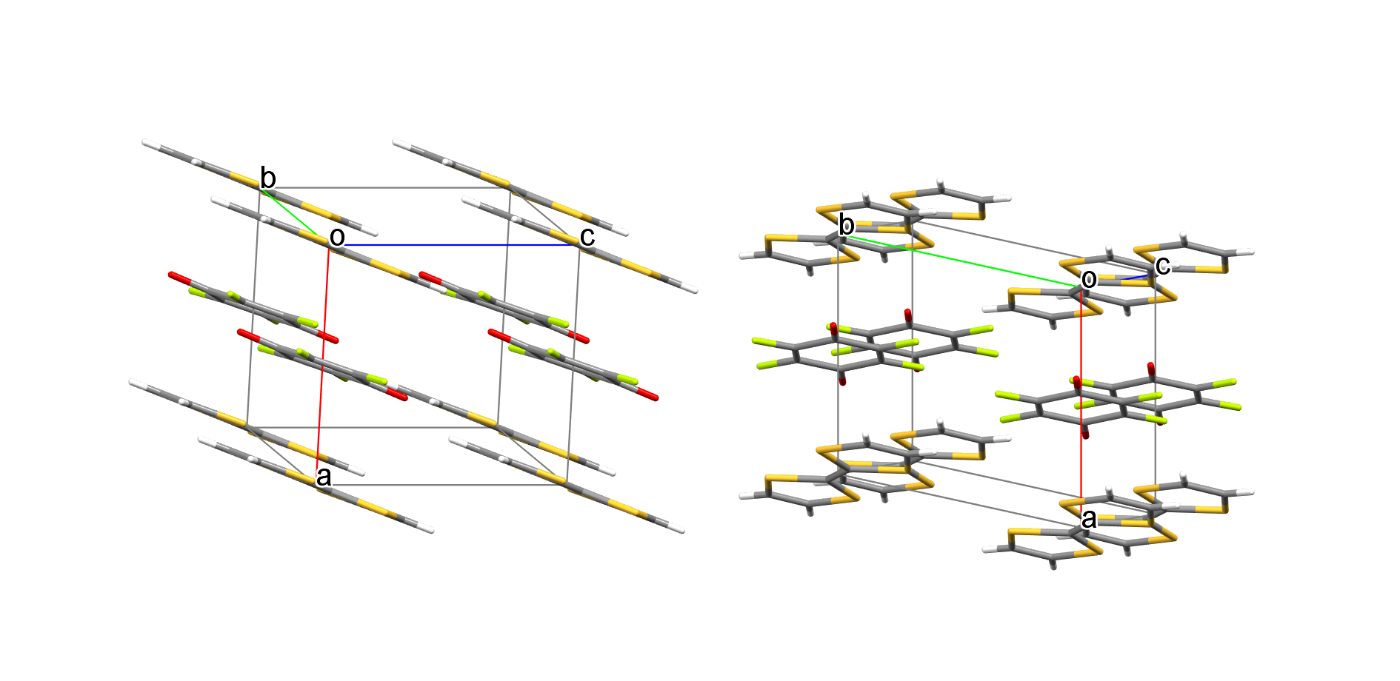}

        \caption{Crystal structure of TTF-FA at room temperature, viewed along the [010] (left) and [001] (right) directions.
          The spectra have been recorded on these two faces. The C=O $b_{1u}\nu_{10}$ mode can be excited in the ac-face only.}
	\label{fig:structure}
\end{figure}

Single crystal XRD analysis confirmed the triclinic ($P-1$) structure found by Torrance et al.\cite{Mayerle1979},
characterized by a mixed regular stack, with TTF and FA molecules alternating along $a$ and laying on the inversion centers.
The green and needle-shaped TTF-FA crystals (Fig S1) are elongated in the stack direction and usually show two different faces,
parallel to the $ac$ and $ab$ planes (Fig \ref{fig:structure}). The stack direction can be easily recognized by optical spectroscopy due to the presence
of the CT excitation, occurring at 6000 \W (Fig S4). The needle-like crystal shape reflects the strong and directional CT interaction,
that is the main intermolecular force and drives a faster crystal growth along the stack direction.
However, sizeable interstack interactions are present also along the $c$-direction through C-H...F and S...O contacts
between TTF and FA units (Fig S2 and S3).

TTF-FA crystals undergo a temperature induced first order NIT, associated with sample cracking and a sudden colour change from green to brown.
During the first cooling the transition temperature is not well defined and varies in the range 120-160 K depending on sample history and cooling rate.
This is probably due to sample defects and internal strains.
Indeed, the spectra recorded on different points of the same crystal sometimes reveal the coexistence of the two phases\cite{LeCointe2003}.
The transition is reversible with a large thermal hysteresis, with the ionic to neutral (I-N) transformation taking place at 200 K on heating
and the N-I one at 170 K on cooling again, in a reproducible way.
While the optical axes and the polarization direction of the CT excitation are unchanged by the transition,
the strongly discontinuous evolution of the IR and Raman spectra (Fig S6) indicates a large ionicity jump,
stack dimerization and doubling of the unit cell, as described in the following sections.

\subsection{IR spectra: Ionicity} 
The NIT is characterized by two order parameters: the ionicity and the stack dimerization.
Both can be conveniently studied by vibrational spectroscopy: the ionicity affects the vibrational frequencies of D and A
while the dimerization reduces the lattice symmetry.\par
\begin{figure}[hp]
	\centering
       	\includegraphics[width=\linewidth]{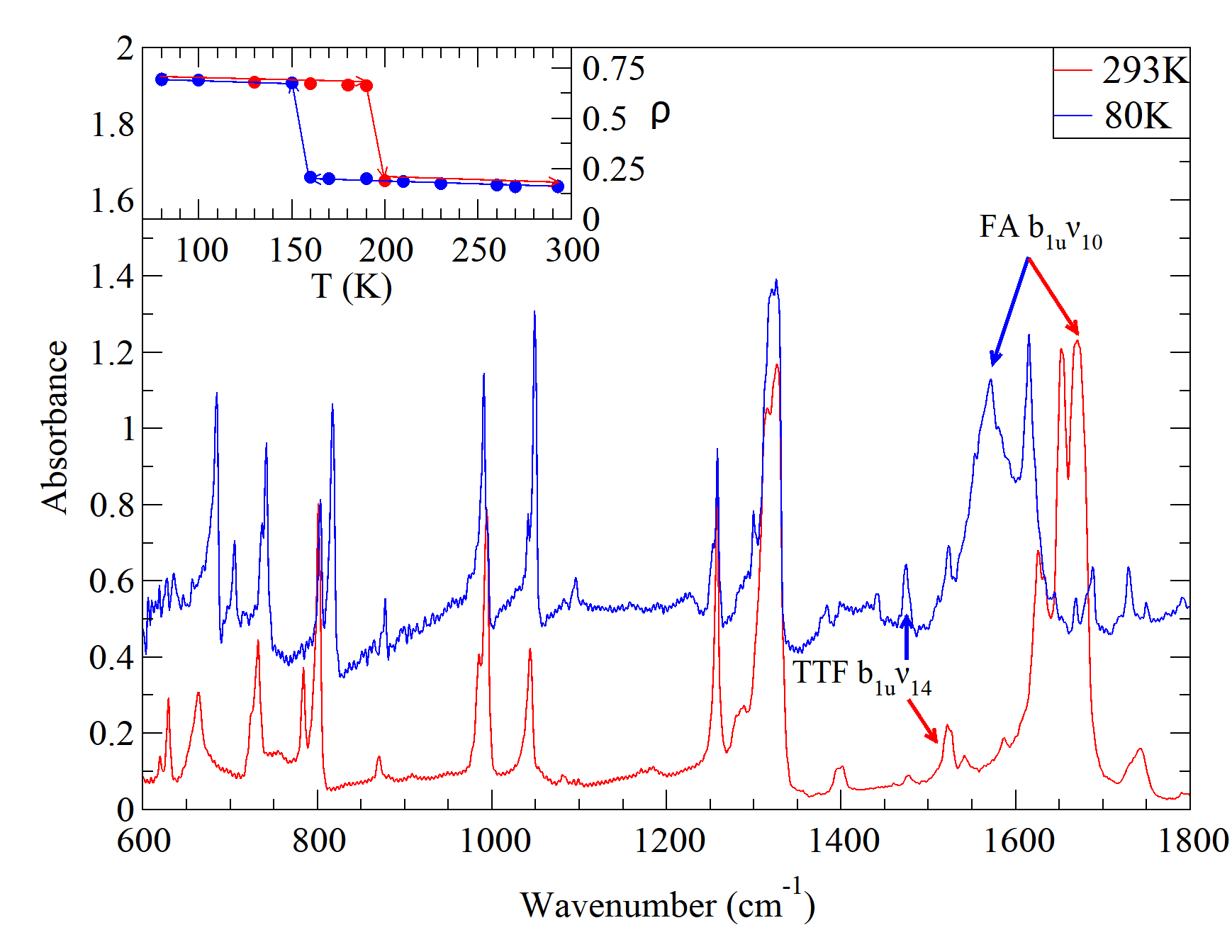}
        \caption{IR spectra of the N- and I- phases of TTF-FA, recorded on $ac$ face with the electric field polarized perpendicular to the stack. The spectra are upshifted for clarity. Inset: Ionicity changes of TTF-FA estimated from the FA $b_{1u}\nu_{10}$ frequency: first cooling (blue) and heating (red). }
	\label{fig:ionicity}
\end{figure}

 
The degree of charge transfer can be obtained from the frequency shifts of some selected “charge sensitive modes” \cite{Ranzieri-JPC2007} of the constituent molecules.
Since the frequencies of the totally symmetric vibrations can be perturbed by e-mv interaction \cite{Girlando-PRB1982,Painelli1986-bibbia},
the most reliable modes are in-plane, antisymmetric with respect to the inversion center and can be observed in the IR spectra
polarized perpendicular to the stack. Since in a first approximation the $D_{2h}$ point group of the TTF and FA molecules is retained in the crystal,
these vibrations, polarized along the two in-plane molecular axes, have $b_{1u}$ and $b_{2u}$ symmetry.
While the spectra recorded on the $ab$ face are dominated by the FA $b_{2u}$ bands (see Fig. \ref{fig:structure} and Fig S5, upper panel),
the spectra recorded on the $ac$ face, reported in Fig \ref{fig:ionicity}, show mainly the $b_{1u}$ vibrations and a small projection of $b_{2u}$ ones.
The complete assignment of the IR spectra is reported in Table S1.\par

\begin{table}[h]
\caption{\label{tab:table1}
Assignment of some selected modes of FA in TTF-FA N and I phases. The frequencies are in \W and the calculated values have been scaled. }
\begin{ruledtabular}
\begin{tabular}{lcccc}
Mode&FA&
TTF-FA 293K&
TTF-FA 80K&
FA$^{-}$\\
 &expt (calc)&
 &
 &
(calc)\\
\hline
$a_{g}\nu_{1}$&1704 (1709)&1642&1547&(1545)\\
$a_{g}\nu_{3}$&1251 (1290)& 1233 & 1170 & (1269) \\
$b_{1u}\nu_{10}$
  &1700 (1704)
  & 1670 & 1572 & (1520) \\
  $b_{2u}\nu_{18}$
  &1669 (1649)
  & 1651 & 1616 & (1532) \\
\end{tabular}
\end{ruledtabular}
\end{table}

\begin{table}[h]
\caption{\label{tab:table2}
Assignment of some selected modes of TTF in TTF-FA N and I phases. The frequencies are in \W and the calculated values have been scaled. }
\begin{ruledtabular}
\begin{tabular}{lcccc}
Mode&TTF&
TTF-FA 293K&
TTF-FA 80K&
TTF$^{+}$\\
 &expt(calc)&
 &
 &
expt(calc)\\
\hline
$a_{g}\nu_{2}$&1555(1582)&1540&1522&1505(1511)\\
$a_{g}\nu_{3}$&1518(1538)& 1475 & 1384 & 1420(1393) \\
$b_{1u}\nu_{14}$
  &1530(1560)
  & 1527 & 1478 & 1478(1492) \\
  $b_{1u}\nu_{16}$
  &781 (782)
  & 785 & 803 & 836 (831)\\
  $b_{3u}\nu_{34}$
  &639 (654)
  & 657 & 680 & 705 (709)\\
\end{tabular}
\end{ruledtabular}
\end{table}

In the spectrum of the high temperature phase the C=O and C=C stretching bands of FA,
named $b_{1u}\nu_{10}$ and $b_{2u}\nu_{18}$ respectively, are redshifted compared to the same bands of the pure compound, as expected \cite{Girlando1975-FA}.
In fact, in passing from haloquinone neutral molecules to their radical anions these double bonds acquire partially single bond character
and for this reason their stretching frequencies are strongly lowered \cite{Girlando1978}. Furthermore, their frequency shifts are known to show a linear behavior,
making both modes good ionicity probes: $b_{1u}\nu_{10}$ in the whole ionicity range and $b_{2u}\nu_{18}$ for neutral ($\rho<0.5$) compounds only \cite{Ranzieri-JPC2007}.
In the case of FA the ionization frequency shifts were estimated through DFT calculations due to the lack of experimental data on FA anion.
According to the calculations, $b_{1u}\nu_{10}$ and $b_{2u}\nu_{18}$ are redshifted 184 and 117 \W, respectively (Table \ref{tab:table1}).
The $b_{1u}\nu_{10}$ frequency full ionicity shift is consistent with that of other haloquinones, which is 160 \W for both chloranil and bromanil \cite{Girlando1978}.
At room temperature $b_{1u}\nu_{10}$ and $b_{2u}\nu_{18}$ occur at 1670 and 1651 \W, indicating that TTF-FA is on the neutral side with a degree of charge transfer of 0.15.
Also the frequency shifts of other selected TTF charge sensitive modes \cite{Bozio1979}, $b_{1u}\nu_{14}$, $b_{1u}\nu_{16}$ and $b_{3u}\nu_{34}$ (Table \ref{tab:table2}),
yielding values in the range 0.1-0.3, confirm the above estimate. \par
After the phase transition all the charge sensitive bands of TTF and FA shift, showing a large ionicity increase (Fig \ref{fig:ionicity}).
In particular, the two strong bands observed at 1572 and 1616 \W can be assigned to FA $b_{1u}\nu_{10}$ and $b_{2u}\nu_{18}$, respectively.
In fact, the former band is present only in the spectrum recorded on the ac face, while the latter is found in the spectra of both $ac$ and $ab$ faces,
as for the high temperature phase (Fig S5, lower panel).
The $b_{1u}\nu_{10}$ frequency corresponds to a ionicity value of 0.7, showing that TTF-FA crosses the neutral-ionic borderline at the transition.
The temperature evolution of $\rho$ estimated from the $b_{1u}\nu_{10}$ mode, shown in the inset of Fig \ref{fig:ionicity}, is strongly discontinuous,
with very little changes on cooling above and below the transition.
In the case of TTF, the antisymmetric C=C stretching band is shifted to 1478 \W, as in the completely ionized molecules ($\rho=1$).
However, the other charge sensitive modes indicate a lower degree of charge transfer (0.6 for $b_{3u}\nu_{34}$ and 0.4 for $b_{1u}\nu_{16}$).
Also the frequency shifts of two Raman active TTF totally symmetric modes, $a_{g}\nu_{2}$ and $\nu_{3}$, yield a ionicity value of 0.6-0.7,
if the effects of e-mv coupling are considered\cite{Painelli1986-bibbia,Masino2017-Crystals}, assuming the CT transition frequency to be 6000 \W as in the neutral phase.
The small discrepancies between the TTF bands are probably due to the deformation of the molecules in the ionic phase, as described in the following.

\subsection{Vibronic effects: stack dimerization}

\begin{figure}[hp]
	\centering
       	\includegraphics[width=\linewidth]{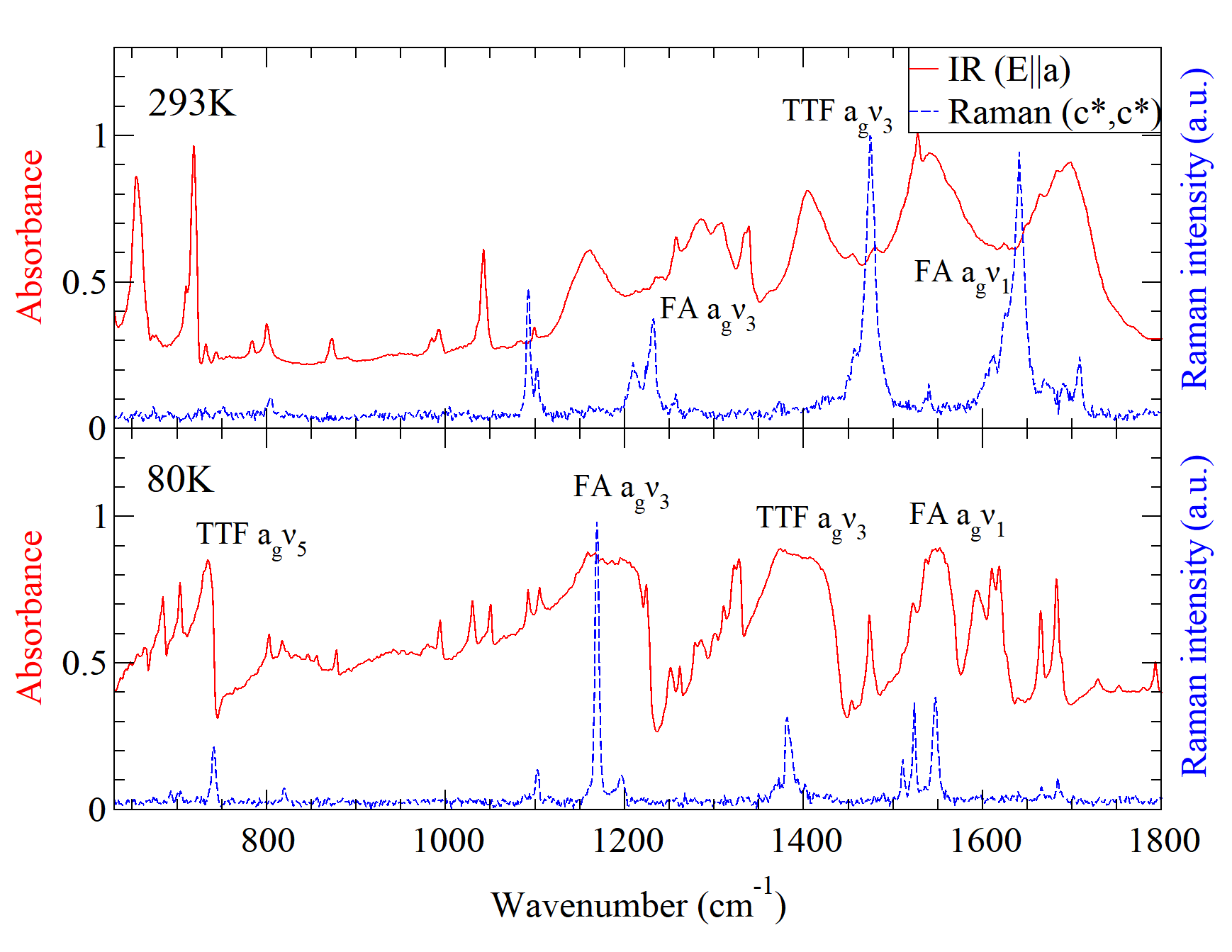}
        \caption{Comparison of TTF-FA parallel polarized IR spectrum and Raman spectrum recorded on $ac$ face with both exciting and scattered light polarized perpendicular to the stack, at 293 K (upper panel) and 80 K (lower panel). $\lambda_{exc}$= 568.2 nm }
	\label{fig:dimerization}
\end{figure}

While the IR spectrum polarized perpendicular to the stack gives information on the ionicity,
the parallel polarized one, being affected by vibronic effects, is very sensitive to stack symmetry \cite{Girlando-PRB1982,Painelli1986-bibbia}.
In the IR spectrum of the neutral phase polarized parallel to the stack (Fig \ref{fig:dimerization}, upper panel)
the absence of vibronic activation of the totally symmetric modes demonstrates that the stack is regular,
with both molecules laying on inversion centers. However, some strong and broad IR absorptions are found at the same distance, about 65 \W, from the most strongly coupled totally symmetric modes, i.e. TTF $a_{g}\nu_{3}$ and FA $a_{g}\nu_{1}$ and $\nu_{2}$. Similar features were observed in the IR spectra of the centrosymmetric neutral phase of TTF-CA \cite{Masino2003}. They are the so-called sidebands, two phonon excitations due to sum and difference combinations between the totally symmetric intramolecular vibrations and an effective Peierls phonon, the lattice mode that drives the stack dimerization. As in the case of TTF-CA, all three pairs of sidebands shift towards the corresponding totally symmetric bands on lowering temperature (Fig S7). This evolution reflects the softening of the Peierls mode that remains incomplete, interrupted by the discontinuous first order transition.\par

Differently, the parallel polarized IR spectrum of the low temperature phase shows saturated absorptions in correspondence with the three most strongly coupled totally symmetric vibrations of TTF and FA, observed in the Raman spectrum at nearly the same frequencies (Fig \ref{fig:dimerization}, lower panel). The IR activation of these vibrations is a proof of the stack dimerization \cite{Painelli1986-bibbia}. Unluckily, it was not possible to follow the temperature evolution of these bands as they saturate the absorption spectrum and the reflectivity measurements were not possible due to the damage of the crystal surface during the transition.\par

\subsection{Lattice phonons: doubling of the unit cell}
Some information on the stack symmetry and on specific lattice vibrations, such as the effective Peierls mode, have been obtained from the parallel polarized IR spectra in the range of the intramolecular vibrations thanks to vibronic effects. A deeper insight on lattice symmetry can be achieved by direct Raman observation of the lattice vibrations in the range 5-200 \W.
These spectra are very sensitive to crystal packing and the number of Raman active intermolecular vibrations can provide information
on the symmetry and multiplicity of the unit cell. \par
\begin{figure}[h]
	\centering
       	\includegraphics[width=\linewidth]{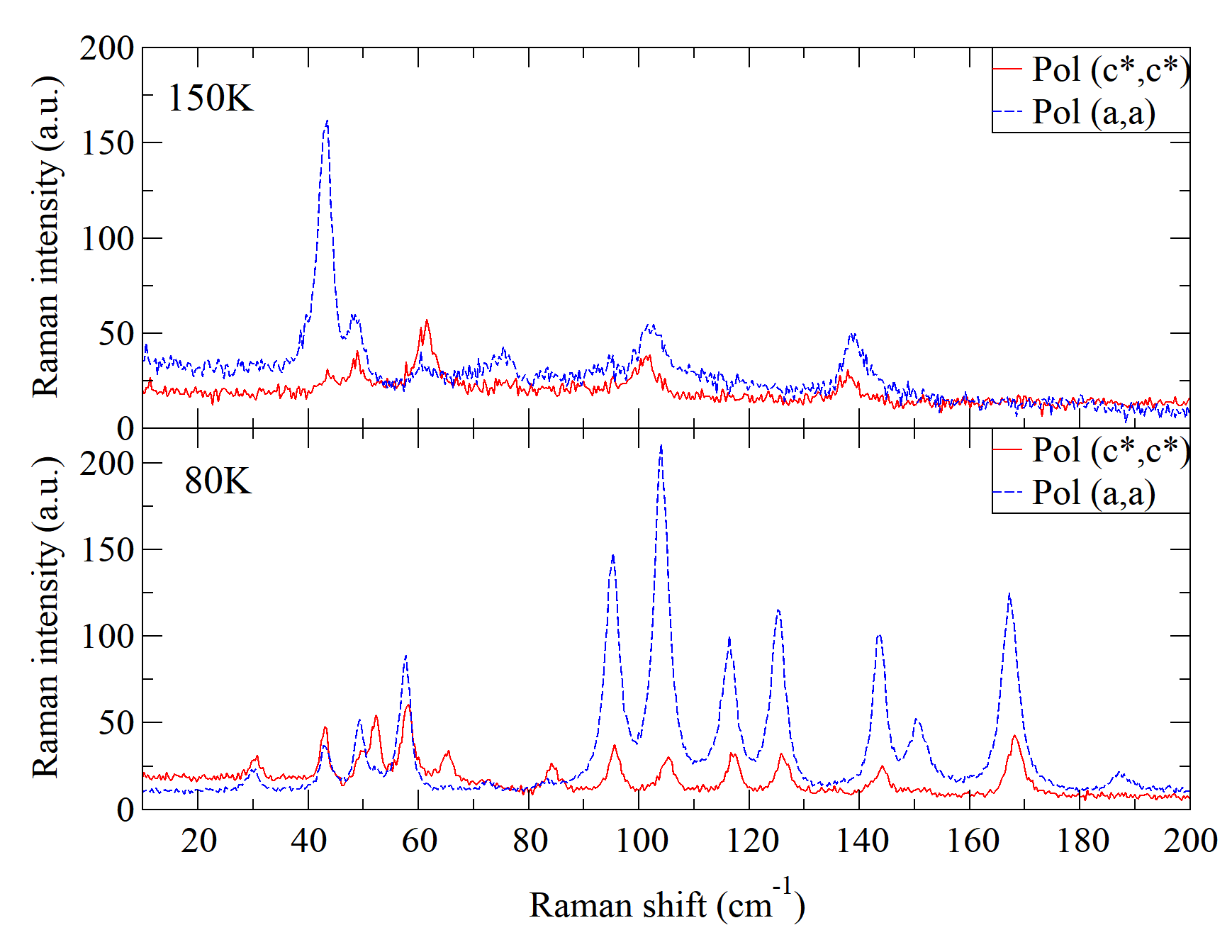}
        \caption{Low frequency Raman spectra of TTF-FA N- (upper panel) and I- phases (lower panel), recorded on $ac$ face. The polarization of exciting and scattered light is the same: perpendicular (red continuous line) or parallel (blue dashed line) to the stack. The relative magnitudes of all the spectra reflect the intensity of the light scattered by the sample (the parallel polarized $(a,a)$ spectrum of the I-phase has been scaled by a factor of 3 for clarity). $\lambda_{exc}$= 633 nm  }
	\label{fig:phonons}
\end{figure}

In the case of the neutral phase, the $P-1$ unit cell contains one D-A pair and both molecules occupy sites with $C_{i}$ symmetry.
Consequently, 9 optically active lattice phonons are expected. Since the mutual exclusion rule holds,
the six rotations should be Raman active while the three translations IR active.
In fact, the low frequency Raman spectra of the neutral phase show six bands (Fig \ref{fig:phonons}, upper panel).
Both the polarizations reported here (spectra recorded on $ac$ face) share the same bands with different relative intensities.
The other polarizations do not add relevant information.\par

The phase transition is revealed by the appearance of many new bands and by the intensity enhancement of the $(a;a)$ polarized spectrum, in particular in the range 80-170 \W (Fig \ref{fig:phonons}, lower panel),
due to the activation of translational modes coupled to the CT excitations.
This spectrum shows 12 bands, most of which are in common with the $(c^{*},c^{*})$ polarization.
Overall, 15 different bands can be distinguished in the spectral range 0-180 \W. Differently, in the $(b^{*},b^{*})$ polarization the majority of these bands are very weak (Fig S8),
suggesting the occurrence of cell doubling along the $c$-axis.
If a triclinic unit cell is doubled, the arrangement of the two adjacent dimerized stacks can be either \textit{ferrielectric}, if the stacks are inequivalent, or \textit{antiferroelectric} 
if the stacks are equivalent and related by inversion symmetry.
In both cases 21 optically active phonons are expected, 
but the number of Raman active ones is different in each situation.
In the former case all the phonons would be both IR and Raman active due to the lack of inversion symmetry.
Differently, in the latter case, due to the presence of inversion center situated between the chains,
the mutual exclusion rule would be still valid and only 12 Raman active phonons (6 rotations and 6 translations) are predicted.\par
Considering the fact that at 80 K the increased resolution due to band narrowing makes the weakest bands detectable, the ferrielectric arrangement is unlikely. 
In fact, both TTF and FA have low frequency intramolecular vibrations \cite{Girlando1975-FA,Bozio1979}.
All of them might be Raman active due to the symmetry lowering of the molecule in the crystal.
In these circumstances, in absence of a symmetry center more than 21 Raman bands would be expected, in disagreement with the 15 ones observed.

\subsection{Crystal structure of the ionic phase}

Below the transition temperature, single crystal diffraction clearly indicates a 
change of the lattice parameters being now $a$=7.2023(2), $b$=7.8625(3), $c$=12.6272(4) \AA, $\alpha$=83.040(2), $\beta$= 73.717(1), $\gamma$=77.224(2)°.
Structure refinement in $P-1$ gave R1 agreement factor of 3.30 (Table \ref{tab:table3}), no relevant improvement was observed in the $P1$ symmetry,
despite doubling of the parameters, as a consequence the centrosymmetric solution was chosen.
Respect to the spatial arrangement of the molecules, the $a$ and $b$ axes of the neutral phase respectively correspond to the $b$ and $a$ axes of the ionic phase,
due to a crystallographic convention. In the following, all the axis labels are referred to the neutral phase. \par
\begin{figure}[h]
	\centering
       	\includegraphics[width=\linewidth]{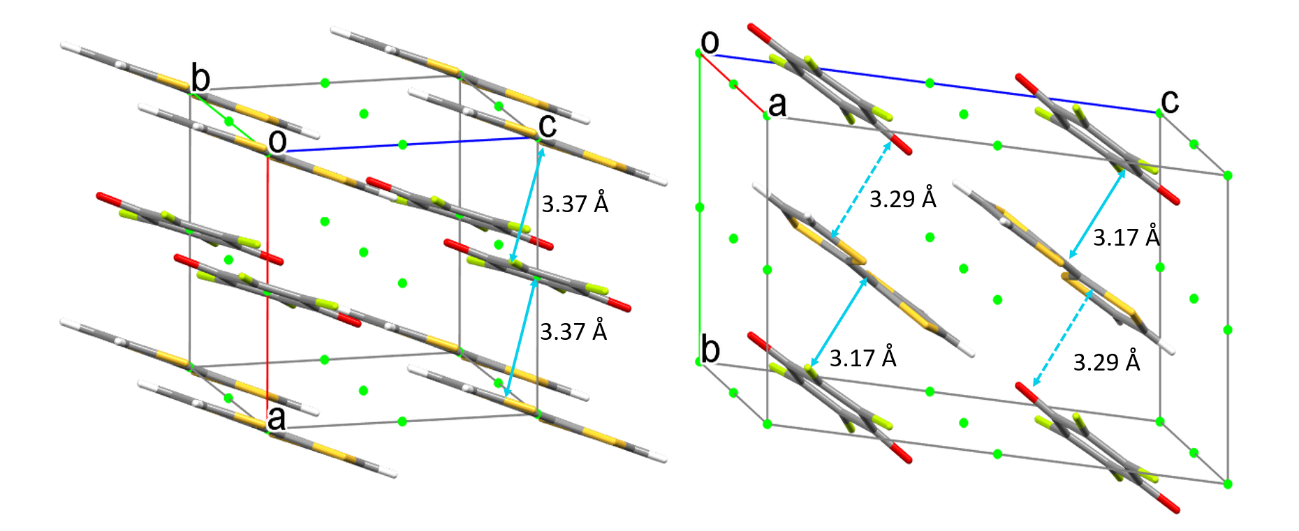}
        \caption{Crystal structures of TTF-FA at 293 K (left) and 100 K (right), viewed along the [010] direction. The inversion centers are marked in green. Intra- and inter-dimer D-A distances (molecular plane to centroid) are reported. }
	\label{fig:LTstructure}
\end{figure}

The ionic phase is characterized by the doubling of the unit cell along the $c$-axis, with two equivalent stacks, strongly dimerized and related by inversion symmetry in an \textit{antiferroelectric} 
arrangement (Fig \ref{fig:LTstructure}, right).
The intradimer and interdimer distances (plane-centroid) are 3.17 and 3.29 Å, respectively.
The intradimer relative orientation of TTF and FA is similar to that observed in the neutral phase.
On the contrary, the interdimer interaction is reduced as the dimers are shifted respect to each other approximately along the $c$-axis.
This increases the angle between the stack axis and the FA molecular planes from 18.3° to 35.3°.
For the same reason the $\beta$ angle changes from 93.1° to 83.0° and the a-axis increases from 7.054 to 7.863 \AA (Table \ref{tab:table3}).
Furthermore, the antiferroelectric arrangement of the ionic phase might be related to the interchain interactions
that are also present in the neutral phase (Fig S2 and S3). This motif is maintained in the ionic phase.
Finally, we noticed that TTF molecules, no longer laying on the inversion centers, are not planar, being twisted around the central C=C bond.
This is probably a consequence of the double bond weakening due to ionization.  

\begin{table}[h]
\caption{\label{tab:table3}
Structural parameters of N and I phases of TTF-FA. {*}Respect to the spatial arrangement of the molecules, the a and b axes of the N phase respectively correspond to the b and a axes of the I phase, due to a crystallographic convention.}
\begin{ruledtabular}
\begin{tabular}{lcc}
 &Neutral phase (293K)&
Ionic phase (100K)$^{*}$\\
\hline
Crystal system&Triclinic&Triclinic\\
Space group&$P-1$&$P-1$\\
$a$ (Å) &7.054(5)& 7.2023(2) \\
$b$ (Å)&7.304(6)&7.8625(3) \\
$c$ (Å) &7.359(5) & 12.6272(4) \\
$\alpha$(deg) &106.150(12)&	83.040(2) \\
$\beta$(deg) &93.101(14)&	73.7170(10) \\
$\gamma$(deg) &102.419(14)&	77.224(2) \\
Z &1 & 2 \\
V (Å$^{3}$) &353.0(5) &	668.04 \\
R	&4.03&	3.30 \\
\end{tabular}
\end{ruledtabular}
\end{table}

\section{Conclusions}

In this study the temperature induced Neutral to Ionic phase transition of TTF-FA has been investigated by vibrational spectroscopy and single crystal XRD.
The transition is first order and associated with a large ionicity jump ($\Delta\rho\simeq 0.5$), stack dimerization and doubling of the unit cell,
taking place suddenly and simultaneously, together with an explosive sample cracking.
The antiferroelectric arrangement of the chains in the ionic phase is the most likely from both spectroscopic and structural data.
Also the absence of splitting of all the charge sensitive bands suggests the equivalence of the two stacks. \par
In each phase the temperature evolution of the IR and Raman spectra is strongly discontinuous
and does not anticipate the corresponding spectra of the other phase approaching the transition.
The same behavior is common with the first order NIT observed in TTF-CA,
but with a stronger discontinuity between the spectra of the two phases and a much larger thermal hysteresis.
However, the space groups of the two CT systems are different: the TTF-CA N-phase is monoclinic $P_{21}/n$,
with already two stacks in the unit cell that dimerize at the transition in the polar $P_{n}$ space group\cite{LeCointe1995}.
On the contrary, TTF-FA N-phase is triclinic with one D-A pair in the unit cell.
At the transition the stacks dimerize in antiphase resulting in an antiferroelectric order, associated with the doubling of the unit cell.
From this perspective TTF-FA broken symmetry is analogous to that of dimethyl-TTF-CA \cite{Ranzieri-PRB2007}.
However, in the latter system the phase transition is close to second order and probably dominated by 1-d stack dimerization over that of charge instability.
In fact, the softening of the effective Peierls mode is almost complete and the ionicity of the low temperature phase remains below 0.5 \cite{Ranzieri-PRB2007}.\par 
Calculations along the lines of Ref. \cite{Delchiaro2017} based on the modified Hubbard model, performed on the crystal structures,
predict instability in a large ionicity range for TTF-FA (Fig S9 and Table S2). This agrees with the strong ionicity jump observed.
Moreover, the computed model parameters allow the comparison between TTF-FA and other ms CT systems.
In this way TTF-FA is close to TTF-CA: despite the differences in crystal packing and D-A overlap,
the two systems share similar microscopic parameters leading to a discontinuous behavior. \par 
Since many simultaneous processes usually play a role in the NIT, the mechanism is often difficult to identify in detail.
However, a quite plausible scenario can be put forward on the basis of the general understanding gained through
the analysis of the extended Peierls-Hubbard model \cite{Painelli1988-ZeroT,DAvino-Crystals2017}. NIT is in fact characterized by the interplay
between the Peierls electron-phonon coupling, yielding the dimerization of the  stack, and the Madelung energy which causes the $\rho$ increase.
In this case it is clear that the Madelung energy increase following thermal lattice contraction prevails,
inducing the observed  discontinuous  ionicity variation.
Furthermore, the low steric hindrance of FA, the smallest haloquinone,
reduces the D-A intermolecular distances increasing 3-d interactions and the Madelung energy.
This is the most likely explanation for the relatively high transition temperature of TTF-FA,
compared to other TTF-haloquinone complexes \cite{Horiuchi2001-jacs,Horiuchi2006-ChemPhys}.
Also, the sizeable interstack interactions between TTF and FA certainly play an important role in the doubling of the unit cell,
leading to an antiferroelectric ordering. Finally,the presence of metastable phases and the large thermal hysteresis suggest a high activation energy, probably due to the drastic structural rearrangement during the transition.  \par
Interestingly, systems showing discontinuous NIT with a high activation barrier might be subject to photoinduced NIT as well. This means that a pulsed excitation can potentially generate a long-lived metastable phase, as in the case of TTF-CA \cite{Koshihara1999}. In addition, it is well known that the few mixed stack CT showing temperature induced NIT undergo the transition at a lower pressure ($\simeq1$GPa),
compared to other systems undergoing pressure induced NIT only \cite{Horiuchi2006-ChemPhys}.
Thus, TTF-FA is expected to show NIT upon the application of a modest pressure, as suggested in Ref. \cite{Torrance1981}.
Further studies are needed to confirm this hypothesis and to compare the results with those of the widely studied TTF-CA \cite{Hanfland1988}, thus achieving a deeper understanding of the transition.\par

\section{Acknowledgements}
This project benefited from the equipment and support of the COMP-HUB Initiative, funded by the “Departments of Excellence” program of the Italian Ministry for Education, University and Research (MIUR, 2018-2022).
The authors gratefully acknowledge Prof A. Girlando for the enlightening discussions.

\bibliography{TTF-FA}

\end{document}